\documentclass[lettersize,journal]{IEEEtran}
\usepackage{amsmath,amsfonts}
\usepackage{algorithmic}
\usepackage{algorithm}
\usepackage{array}
\usepackage{hyperref}
\usepackage[caption=false,font=normalsize,labelfont=sf,textfont=sf]{subfig}
\usepackage{booktabs} 
\usepackage{lineno,hyperref}
\usepackage{textcomp}
\usepackage{stfloats}
\usepackage{url}
\usepackage{caption}
\usepackage{verbatim}
\usepackage{graphicx}
\usepackage{cite}
\usepackage{microtype}    
\usepackage{xcolor}     
\usepackage{graphicx} 
\usepackage{amsmath}
\usepackage{algorithm}
\usepackage{algorithmic}
\usepackage{enumitem}
\usepackage{lipsum} 
\usepackage{caption}
\usepackage{subcaption}
\usepackage{multicol}
\usepackage[numbers]{natbib}
\usepackage{comment}
\renewcommand{\textcolor}[2]{#2}
\begin{document}

\title{High Frequency Matters: Uncertainty Guided Image Compression with Wavelet Diffusion}

\author{Juan Song, Jiaxiang He, Lijie Yang$^\ddagger$, Mingtao Feng, and Keyan Wang. 

\thanks{Juan Song, Jiaxiang He, Lijie Yang, Mingtao Feng, and Keyan Wang are with Xidian University, Xi’an 710071, China (Email: songjuan@mail.xidian.edu.cn; 
hjx1255216006@163.com;  
23031212033@stu.xidian.edu.cn; 
mintfeng@hnu.edu.cn; 
kywang@mail.xidian.edu.cn.
)} 


\thanks{$^\ddagger$ denotes corresponding authors.}

\thanks{Manuscript received 2024; revised 2024.}

}

\markboth{Journal of \LaTeX\ Class Files,~Vol.~14, No.~8, 2024}%
{Shell \MakeLowercase{\textit{et al.}}: A Sample Article Using IEEEtran.cls for IEEE Journals}

\maketitle

\begin{abstract}

Diffusion probabilistic models have recently achieved remarkable success in generating high-quality images. However, balancing high perceptual quality and low distortion remains challenging in application of diffusion models in image compression. To address this issue, we propose a novel Uncertainty-Guided image compression approach with wavelet Diffusion (UGDiff). Our approach focuses on high frequency compression via the wavelet transform, since high frequency components are crucial for reconstructing image details. We introduce a wavelet conditional diffusion model for high frequency prediction, followed by a residual codec that compresses and transmits prediction residuals to the decoder. This diffusion prediction-then-residual compression paradigm effectively addresses the low fidelity issue common in direct reconstructions by existing diffusion models. Considering the uncertainty from the random sampling of the diffusion model, we further design an uncertainty-weighted rate-distortion (R-D) loss tailored for residual compression, providing a more rational trade-off between rate and distortion. Comprehensive experiments on two benchmark datasets validate the effectiveness of UGDiff, surpassing state-of-the-art image compression methods in R-D performance, perceptual quality, subjective quality, and inference time. Our code is available at: \href{https://github.com/hejiaxiang1/Wavelet-Diffusion/tree/main}{https://github.com/hejiaxiang1/Wavelet-Diffusion/tree/main}.

\end{abstract}

\vspace{-1mm}

\vspace{-2mm}
\section{Introduction}

\begin{figure}[t]
\centering
\includegraphics[width=1\linewidth]{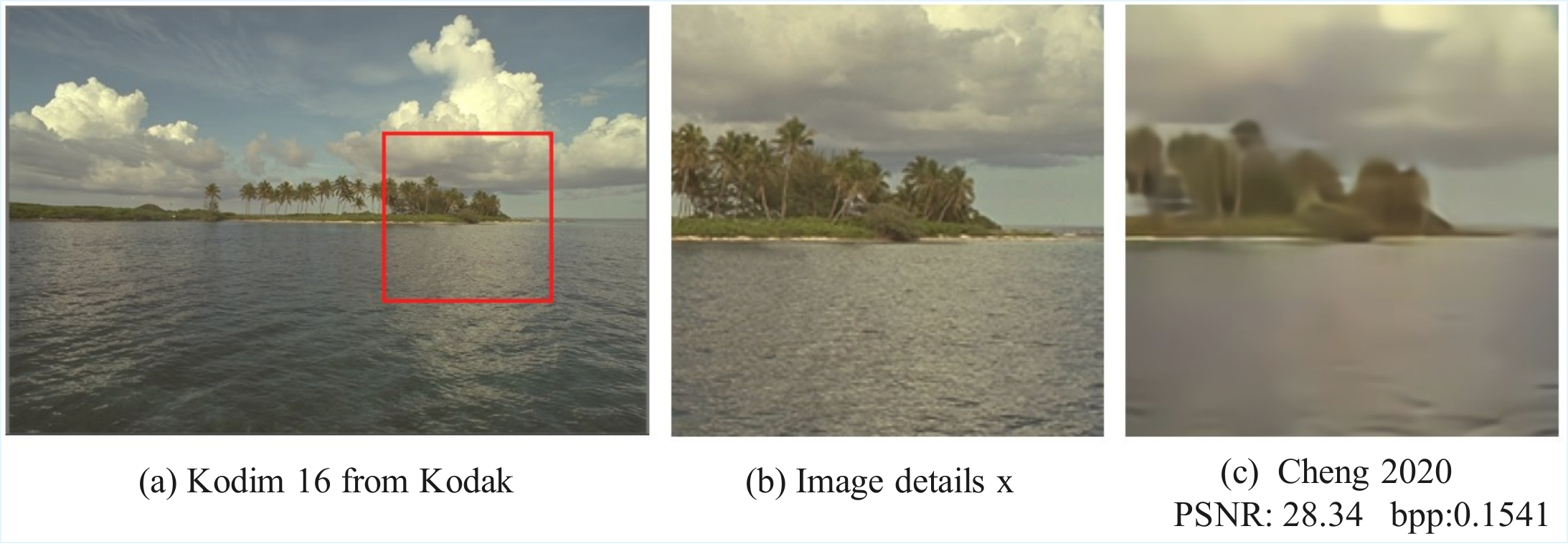} 
\vspace{-5mm}
\caption{Illustration of image details (b) and image reconstructed by an end-to-end learned image compression network(cheng2020~\cite{cheng2020learned}) (c).}
\label{fig:difference}
\vspace{-7mm}
\end{figure}

\IEEEPARstart{I}{n} the era of explosive growth in visual media, efficient lossy image compression has become indispensable to reduce storage costs and transmission bandwidth. 
\textcolor{blue}{However, a long-standing challenge in lossy compression lies in trade-off between achieving low distortion, typically measured by metrics such as Mean Squared Error (MSE), and maintaining high perceptual quality. Particularly, high-frequency components of images, such as textures and edges, tend to be perceptually critical yet highly sensitive to quantization. }
As illustrated in Fig.~\ref{fig:difference},  visually salient high frequency details often exhibit significant degradation, even when traditional distortion metrics indicate good performance. This mismatch highlights importance of high frequency in image compression to  preserve perceptual quality without compromising distortion metrics.

Conventional image compression standards, such as JPEG~\cite{wallace1992jpeg}, BPG~\cite{bpg}, and VVC~\cite{bross2021overview}, adopt a hand-crafted pipeline consisting of transform, quantization, and entropy coding. Although widely adopted, these methods are limited in adaptability to diverse image content due to separately hand-crafted modules. Recently, learned image compression techniques based on Variational Auto-Encoders (VAEs)~\cite{2014Auto} have shown improvements in rate-distortion (R-D) performance~\cite{balle2018variational,minnen2018joint,cheng2020learned,liu2023learned}. Despite their success, most VAE-based methods optimize for MSE loss, which often results in over-smoothed reconstructions and loss of visually important details.


Recent works have introduced generative models to enhance perceptual quality. For instance, Generative Adversarial Networks (GANs)~\cite{goodfellow2014generative} have been employed to generate visually plausible textures~\cite{agustsson2023multi,mentzer2020high}. Diffusion models have further advanced  perceptual quality in image restoration tasks by leveraging great generation capacities, including super-resolution~\cite{shang2024resdiff}, low-light enhancement~\cite{lu2023speed}, and inpainting~\cite{lugmayr2022repaint}. Inspired by these successes, diffusion has recently been introduced to image compression~\cite{yang2024lossy,ghouse2023residual,2025-cvpr-MRIDC,2025-AAAI-controllable}. While these diffusion based image compression approaches excel in synthesizing visually rich reconstructions, they often suffer from pixel-level fidelity degradation. 
Moreover, the uncertainty due to the inherent randomness of the denoising process, which begins with Gaussian noise sampling, reveals the instability of reconstructed pixels. 
Nowadays, few works have explored the impact of uncertainty in diffusion sampling on compression effectiveness although diffusion models have achieved success in image compression field.

The central challenge in balancing low distortion and high perceptual quality lies in reconstruction of high frequency details. Enhancing high frequency reconstruction is a nontrivial task because high frequency typically possesses less energy and are therefore more susceptible to distortion compared to low frequency.
\textcolor{blue}{Motivated by these issues, we propose an Uncertainty-Guided image compression approach with wavelet Diffusion (UGDiff) to maintain high perceptual quality as well as low distortion.We leverage discrete wavelet transform (DWT) ~\cite{mallat1999wavelet} to decouple the image into low-frequency and high-frequency components, enabling a dedicated diffusion model for high frequency to predict fine details and a deterministic low-frequency codec to preserve global structure.}

To faithfully reconstruct perceptually critical high-frequency details,we propose a wavelet diffusion based predictive coding framework for high frequency components. Specially, we propose a wavelet diffusion model to predict high frequency, followed by a residual codec that compresses and transmits the prediction residuals. 
\textcolor{blue}{This diffusion prediction-then-residual compression paradigm offers three advantages. Firstly, our framework explicitly transmits the prediction error (i.e., the residual) to the decoder. This ensures that any deviation from the original contents in diffusion will be corrected, effectively resolving the low fidelity issue common in direct reconstructions by existing diffusion models. Secondly, DWT provides a sparser representation that is easier for a network to learn compared to the pixel domain~\cite{moser2023waving}. Finnaly, DWT reduces the image’s spatial size by a factor of four, in accordance with the Nyquist rule~\cite{landau1967sampling}, thereby expediting the inference speed of the denoising function.
To strictly constrain the conditional diffusion to  generate realistic high frequency, we design a low-to-high frequency translator to generate synthetic high frequency from the reconstructed low frequency by leveraging the inter-band relations between low and high frequency components. }

\textcolor{blue}{Despite their impressive generative capabilities, diffusion models exhibit inherent aleatoric uncertainty due to stochastic sampling from Gaussian noise during the reverse denoising process. This uncertainty often manifests as low-quality samples or artifacts. In our predictive coding framework, this uncertainty of diffusion prediction directly governs the magnitude of the prediction residuals, thus necessitating an uncertainty-guided R-D optimization.
Standard MSE treats all high-frequency residuals equally and lacks an explicit mechanism for adaptive bit allocation, which can lead the network to sacrifice perceptually critical regions to minimize overall distortion.
Motivated by this issues, we design an uncertainty-weighted rate-distortion (R-D) loss for residual compression. Specifically, we first estimate an aleatoric uncertainty map of the predicted high-frequency along the reverse diffusion process via Last Layer Laplace
Approximation(LLLA)~\cite{kou2023bayesdiff}. In addition to the hyper-parameter $\lambda$ that balances the overall R-D level, we then introduce an uncertainty-related weight to the distortion terms to prioritize residuals with high uncertainty where human visual sensitivity is high.}
The main contributions are:

\begin{itemize}[left=1em]
\item We propose a wavelet diffusion based predictive coding for high frequency. 
As far as we know, it is the first endeavor to utilize  diffusion prediction-then-residual compression paradigm to maintain the balance between low distortion and high perception quality.
In addition, the combination of DWT and the diffusion model greatly expedites the inference of diffusion model. 
    
\item We introduce a novel uncertainty guided residual compression module, in which an uncertainty weighted R-D loss is designed to prioritize residuals with high uncertainty and allocate more bits to them. Our proposed uncertainty weighted R-D loss provides content adaptive trade-off between rate and distortion.

\item Extensive experiments on two benchmark datasets demonstrate that our UGDiff effectively balances distortion metrics and perceptual quality compared to previous diffusion methods, while achieving significant speedup.
\end{itemize}

\section{Related Work}
\noindent {\bf Learned Image Compression.} 
Learned image compression has achieved significant progress in network architectures and entropy modeling. Ballé et al.~\cite{balle2022end} first proposed an end-to-end compression framework and later introduced a hyperprior to capture spatial dependencies in latent representations~\cite{balle2018variational}. 
Jiang et al.~\cite{jiang2024llic} enhanced spatial feature extraction by integrating adaptive weights with large-receptive-field transformations. Guo et al.~\cite{guo2023enhanced} leveraged enhanced context mining and Transformer-based networks to reduce contextual redundancy and model long-range dependencies.
\textcolor{blue}{Li \textit{et al.}
\cite{2024-ICLR-FTIC}developed a frequency-aware transformer for learned image compression that leverages directional window attention and frequency modulation to improve rate-distortion efficiency.
Fu \textit{et al.}  
\cite{ECCV-2024-weconvene}introduced WeConv and WeChARM, two wavelet-domain modules, significantly boosting LIC performance while maintaining low computational complexity.
Lee et al. 
\cite{ICML-2024-TACO} introduced TACO, a text-adaptive neural image compression framework that injects CLIP-based semantic cues into the encoder via cross-modal attention and joint image-text loss, enabling simultaneous gains in perceptual fidelity and PSNR at standard bitrates. However, it relies on the quality of text generated by humans or machines, and does not take the cost of generating the text into account.}

\noindent {\bf Diffusion Models for Image Compression.}
Diffusion models~\cite{ho2020denoising} have rapidly achieved SOTA performance in image restoration tasks such as super-resolution~\cite{shang2024resdiff} and inpainting~\cite{lugmayr2022repaint}, and have recently been explored for image compression. Yang et al.~\cite{yang2024lossy} replaced the decoder with a conditional diffusion model, while other works~\cite{ghouse2023residual,hoogeboom2023high} first optimized VAE with rate–distortion objective and then applied diffusion models to enhance perceptual quality. Li et al.~\cite{2024Towards} proposed a two-stage extreme compression framework combining VAEs with pre-trained diffusion models, and Pan et al.~\cite{pan2022extreme} introduced text embeddings to guide diffusion-based reconstruction.
However, existing diffusion-based compression methods primarily apply diffusion at the decoder and favor perceptual quality over reconstruction fidelity, resulting in visually pleasing but less faithful outputs. To address this limitation, we propose a diffusion prediction–then–residual compression paradigm that balances perceptual quality and pixel fidelity.

\noindent {\bf Uncertainty in Bayesian Neural Networks. }
Modeling uncertainty in deep learning have improved the performance and robustness of deep networks in many computer vision tasks, including image segmentation~\cite{badrinarayanan2017segnet}, image super-resolution~\cite{ning2021uncertainty} and etc. 
Ning \textit{et al.}
~\cite{ning2021uncertainty} extended uncertainty modeling to image super-resolution, leveraging Bayesian estimation frameworks to model the variance of super-resolution results and achieve more accurate and consistent image enhancement. 
Chan \textit{et al.}
~\cite{chan2024hyper}proposed Hyper-diffusion to accurately estimate epistemic and aleatoric uncertainty of the diffusion model with a single model.
Kou \textit{et al.} proposed BayesDiff that enabled the simultaneous delivery of image samples and pixel-wise uncertainty estimates based on the Last Layer Laplace Approximation (LLLA)~\cite{kou2023bayesdiff}.
Few studies have explored diffusion model uncertainty in image compression. This paper examines its impact on residual compression and proposes an uncertainty-weighted R-D loss for optimization.

\begin{figure*}[t]
\centering
\includegraphics[width=0.95\linewidth]{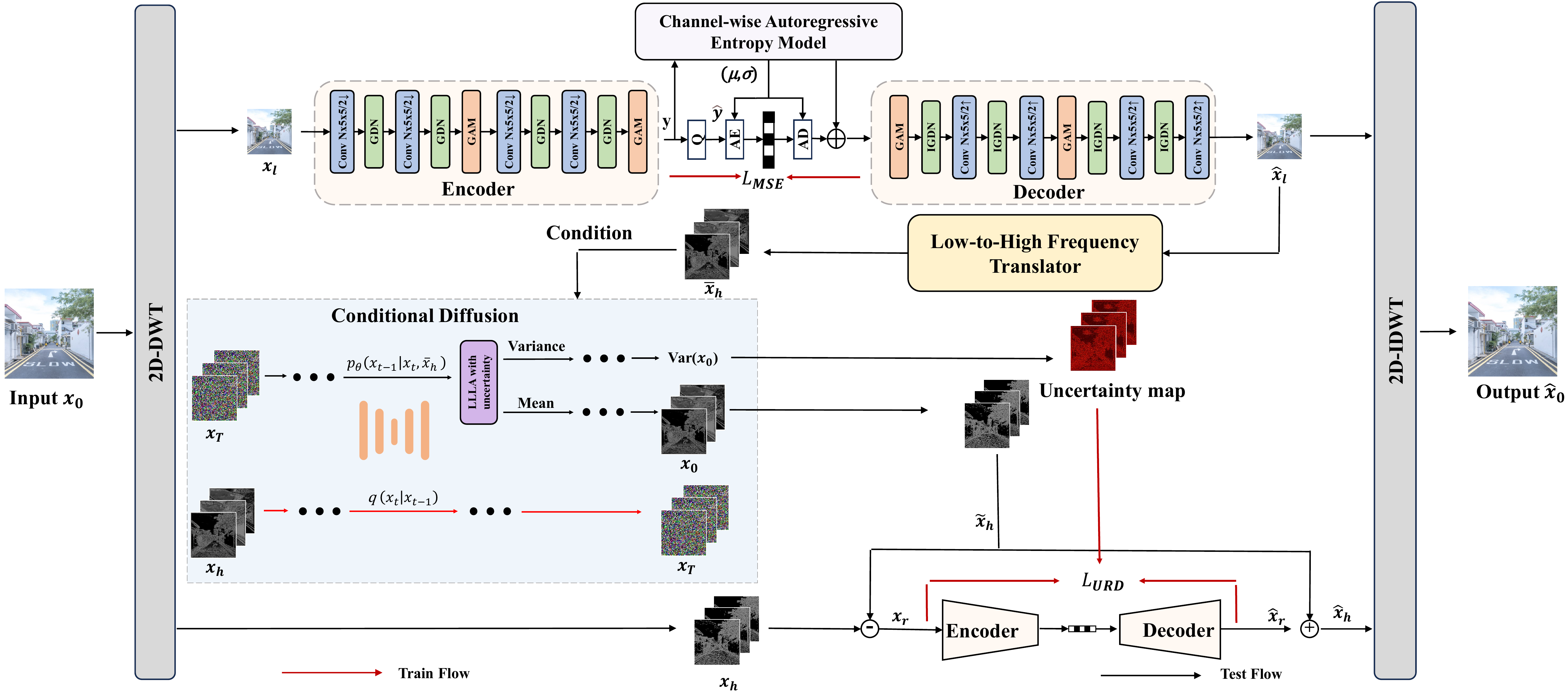}
\caption{Overview of the UGDiff. UGDiff adopts a wavelet diffusion predictive coding pipeline. High frequency is predicted by the conditional diffusion, which is conditioned by synthetic high frequency produced by the low-to-high frequency translator. Simultaneously, the uncertainty map of predicted high frequency is estimated along the reverse diffusion sampling process. The residual between predicted and ground-truth high frequency is then compressed with an uncertainty-weighted residual codec. The reconstructed low- and high-frequency components are finally inversely transformed by 2D-IDWT to reconstruct the image.}
\label{fig:model}
\vspace{-4mm}
\end{figure*}

\section{Proposed Method}
As illustrated in Fig.~\ref{fig:model}, UGDiff adopts a wavelet predictive coding framework, where the input image is first decomposed via DWT and the resulting low and high frequency components are compressed separately. The low-frequency component $x_l$ is encoded using a pre-trained VAE-based codec~\cite{minnen2020channel}. 
Our focus lies on the high frequency branch, where we aim to balance perceptual quality and pixel-level fidelity.
Specifically, a low-to-high frequency translator generates a synthetic high-frequency component $\bar{x}_h$ from the reconstructed low-frequency image. Conditioned by the synthetic high-frequency, a wavelet conditional diffusion model is used to predict rather than directly reconstruct the high frequency. Concurrently, an uncertainty map of the diffusion prediction is estimated via LLLA during reverse diffusion. The residuals $x_r$ between the original and predicted high-frequency components are compressed using a VAE-based codec optimized with an uncertainty-weighted rate–distortion loss, which allocates more bits to regions with higher uncertainty. Finally, the reconstructed image is obtained through 2D-IDWT.

\subsection{Wavelet Conditional Diffusion Model}
\textcolor{blue}
{Directly applying diffusion models to reconstruct images often sacrifices structural fidelity by prioritizing perceptual plausibility, resulting in reconstructed details that diverge from the original content, an outcome that is undesirable for image compression. Moreover, diffusion over the full pixel domain incurs high computational cost due to the large spatial dimension.}

\textcolor{blue}{To address these issues, we propose a wavelet diffusion based predictive coding framework that focuses exclusively on high-frequency components. 
Rather than directly reconstruct high frequency, our approach utilizes the diffusion model to predict high frequency and transmits the prediction residuals to the decoder to correct any deviation
from the original details in diffusion. This diffusion prediction-then-residual compression paradigm effectively mitigates fidelity loss while preserving perceptual quality. Furthermore, by operating on high-frequency subbands that are spatially reduced by a factor of four, our approach dramatically accelerates inference without sacrificing reconstruction accuracy.}

\noindent\textbf{Discrete Wavelet Transform} 
2D-DWT employs a convolutional and sub-sampling operator, denoted as $W$, to transform images from spatial domain to frequency domain, thus enabling the diffusion process solely on high frequency components. 
Let $(x, \hat{x}) \in \mathcal{D}$ denote an original-reconstruction image pair. Before applying the diffusion process, the specific wavelet operator $W$(e.g.haar wavelet), decomposes $x$ into its low frequency component $x_{l}$ and high frequency counterparts $x_{h}$, namely,$({x}_ {l},x_{h}) = Wx$. 

\begin{figure}[t]
\centering
\includegraphics[width=1.0\linewidth]{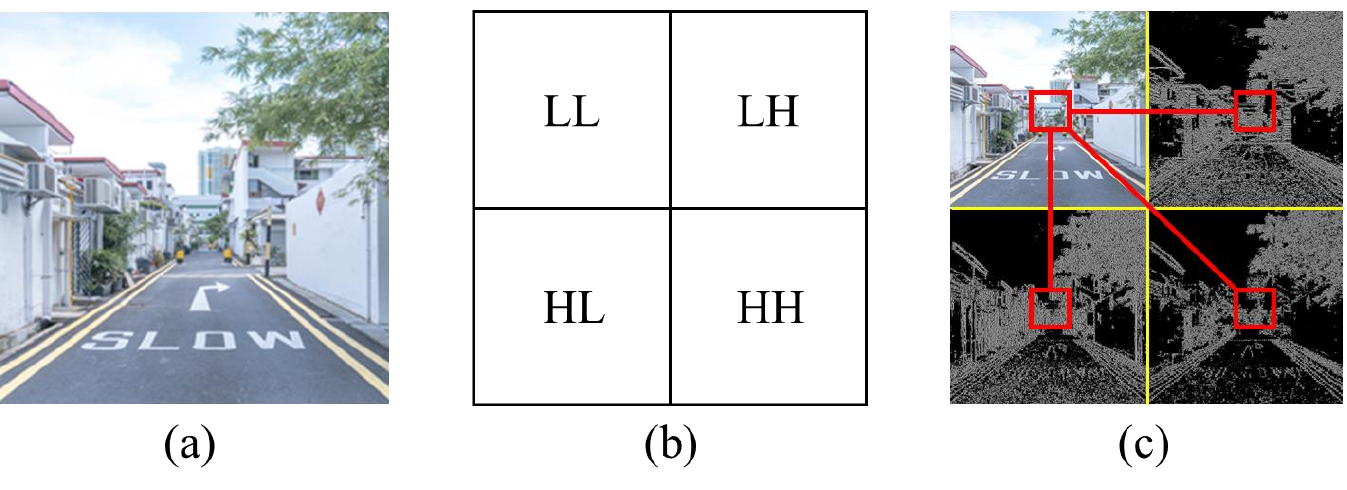}
\vspace{-4mm}
\caption{Wavelet decomposition. (a) Source Image, (b) Wavelet Sub-bands, (c) \textcolor{blue}{Tree structure diagram of the wavelet decomposition. 
There exhibit strong inter-band correlations within the same region (indicated by the red box) sharing similar structure information between low frequency and high frequency components.
}
}
\label{fig:wavelet-decom-abc}
\vspace{-6mm}
\end{figure}

2D-DWT decomposes the image into four sub-bands, namely, Low Low (LL), Low High (LH), High Low (HL) and High High (HH). 
\textcolor{blue}{The low-frequency subband (LL) retains the coarse structural layout of the image, while the high-frequency subbands (LH, HL, HH) encode directional details that are spatially aligned with the LL band,} as illustrated in Fig.~\ref{fig:wavelet-decom-abc}. 


\noindent\textbf{\textcolor{blue}{Low-to-High Frequency Translator.}}
\textcolor{blue}{The aim of our low-to-high frequency translator is not to reconstruct the original high frequency components exactly, but rather to produce a synthetic high-frequency that captures the spatial layout and directional characteristics of the original high-frequency content. This synthetic high-frequency serves as a strong,  structure-aware condition for the subsequent wavelet diffusion model.} 
\textcolor{blue}{In predictive coding paradigm, the decoder has no access to the original high frequency, and reconstructed low-frequency is the only information available at the decoder side.}
A naive approach would condition the diffusion process directly on  reconstructed low frequency $\hat{x}_{l}$,
as the similar way in~\cite{jiang2023low,shang2024resdiff}.
Nonetheless, this is suboptimal as reconstructed low frequency encodes only coarse, low-resolution semantics and lacks high frequency priors needed to guide texture and edge synthesis. Consequently, the diffusion model tends to produce outputs that resemble the smooth structure of low-frequency, failing to recover fine details, as will be illustrated in Fig.\ref{x_l_guide}. 

\begin{figure}[t]
\centering
\includegraphics[width=1.0\linewidth]{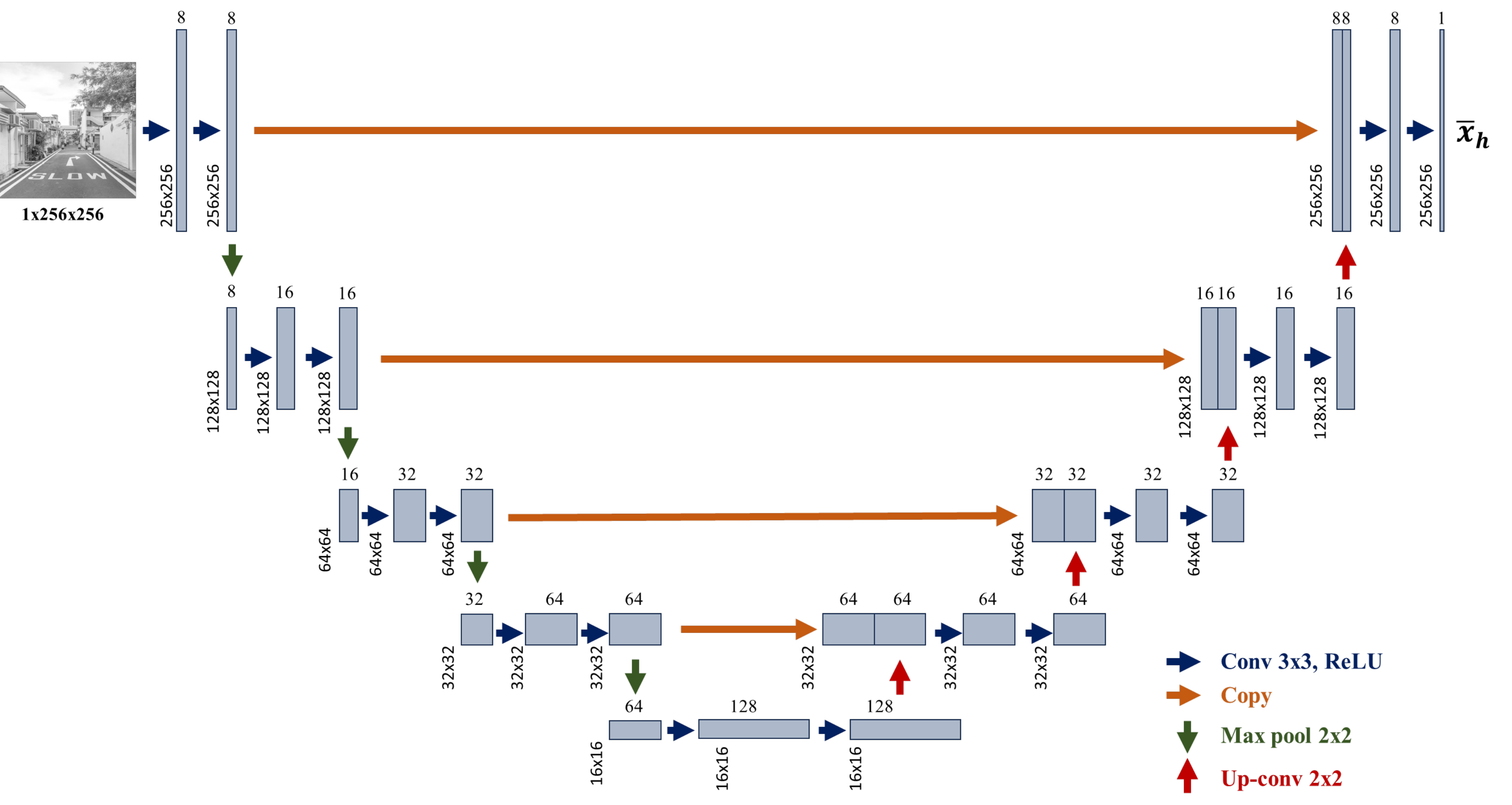}
\vspace{-4mm}
\caption{\textcolor{blue}{Overview of the low-to-high frequency translator.}}
\label{fig:Frequency_conversion}
\vspace{-5mm}
\end{figure}

To derive wavelet diffusion conditions that encapsulate high frequency details from the reconstructed low frequency, we investigate the correlation between the wavelet low frequency and high frequency sub-bands. \textcolor{blue}{As is shown in Fig.\ref{fig:wavelet-decom-abc}, the red boxes highlight strong inter-band correlations, demonstrating that structural information in the low-frequency LL band is spatially aligned with details in the high-frequency bands.
Inspired by the inter-band correlations of wavelet sub-bands, we design a low-to-high frequency translator to map the reconstructed low frequency \textit{$\hat{x}_{l}$} to synthetic high frequency \textit{$\bar{x}_{h}$}, which serves as the condition of the conditional diffusion. }
The frequency translator $G_{\psi}$ is formulated as $\bar{x}_{h}=G_{\psi}\left(\hat{x}_{l}\right)$,
where $G_{\psi}$ is a U-Net-like CNN with localized receptive fields. The network implementation details are shown in Fig\ref{fig:Frequency_conversion}. The encoder consists of four downsampling stages, each with two $3 \times 3$ convolutional layers and a $2 \times 2$ maximum pooling layer, while the decoder mirrors this structure with deconvolution and feature splicing layers.

\noindent\textbf{Conditional Diffusion.}
Equipped with the synthetic high frequency component \textit{$\bar{x}_{h}$} as the condition, we design a wavelet-domain conditional diffusion model in the frequency domain to produce predicted high frequency \textit{$\Tilde{x}_{h}$} with high realism.
A conditional Denoising Diffusion Probabilistic Model (DDPM) utilizes two Markov chains~\cite{ho2020denoising}. The first is a forward chain responsible for adding Gaussian noise to the data:
\begin{equation}
q\left(x_{t} \mid x_{t-1}\right)  =  \mathcal{N}\left(x_{t} ; \sqrt{1-\beta_{t}} x_{t-1}, \beta_{t} I\right)
\end{equation}
where {$\beta_{t}$} represents a variance schedule. 

\begin{figure}[t]
\centering
\includegraphics[width=1.0\linewidth]{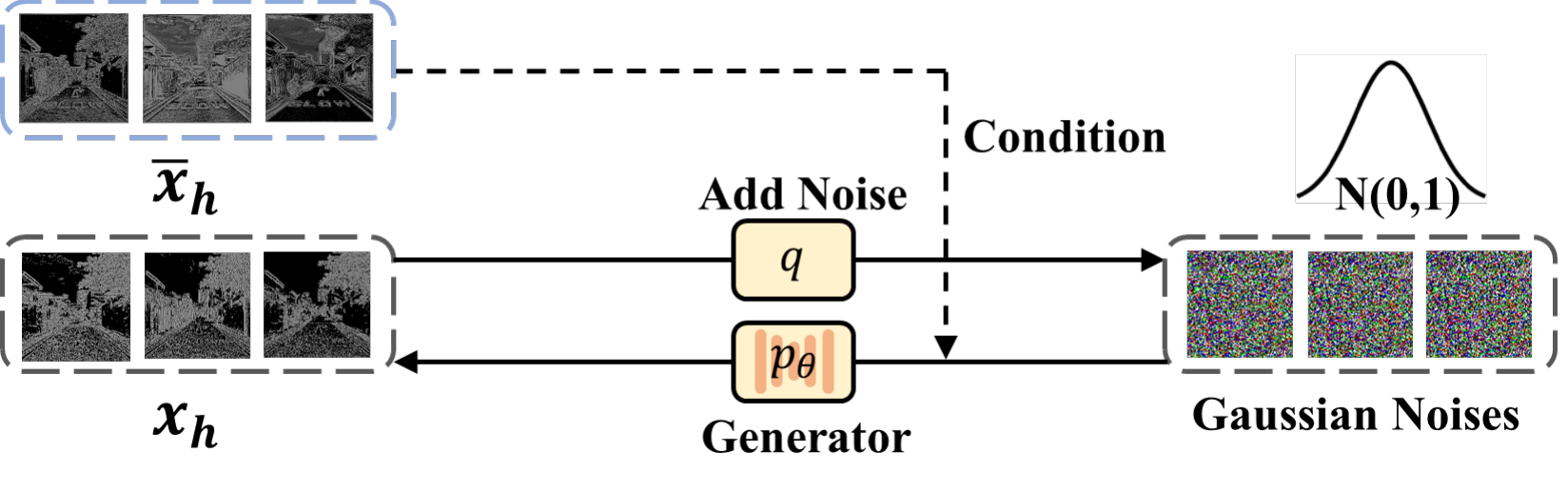}
\vspace{-5mm}
\caption{The forward and reverse process of our conditional diffusion model.}
\label{fig:diffusion-train}
\vspace{-3mm}
\end{figure}

The other Markov chain is a reverse chain that transforms noise back into the original data distribution. As is illustrated in Fig. \ref{fig:diffusion-train}, the key idea of our wavelet conditional diffusion is to introduce the synthetic high frequency \textit{$\bar{x}_{h}$} as the condition into the diffusion model {$\mu_{\theta}\left(x_{t}, t\right)$}, thereby, {$\mu_{\theta}\left(x_{t}, t\right)$} becomes {$\mu_{\theta}\left( x_{t},t,\bar{x}_{h} \right)$}:
\begin{equation}
p_{\theta}\left(x_{t-1} \mid x_{t},\bar{x}_{h}\right)  =  \mathcal{N}\left(x_{t-1} ; \mu_{\theta}\left(x_{t}, t, \bar{x}_{h}\right), \Sigma_{\theta}\right)
\end{equation}
where {$\bar{x}_{h}$} represents conditional guidance that controls the reverse diffusion process. The parameters \textit{$\theta$} are typically optimized by a neural network that predicts {$\mu_{\theta}\left(x_{t}, t, \bar{x}_{h} \right)$} of Gaussian distributions. This is simplified by predicting noise vectors {$\epsilon_{\theta}\left(x_{t}, t, \bar{x}_{h}\right)$} with the following objective:
\begin{equation}
    L_{\text {simple }}=\mathbb{E}_{\mathbf{x}_{0}, t,  {\epsilon}_{t} \sim \mathcal{N}(\mathbf{0}, \mathbf{I})}\left[\left\| {\epsilon}_{t}- {\epsilon}_{\theta}\left(\mathbf{x}_{t}, t,\bar{x}_{h}\right)\right\|^{2}\right]
\label{eq:diffusion_loss}
\end{equation}

\subsection{Uncertainty-guided Residual Compression}



\textcolor{blue}{Diffusion models introduce inherent aleatoric uncertainty due to the randomness of sampling from Gaussian noise, which can lead to unstable prediction residuals in subsequent compression stages. Even within the high-frequency domain, however, not all regions are equally predictable or perceptually significant: complex textures (e.g., hair, foliage) exhibit high uncertainty during diffusion-based prediction, whereas smoother structures (e.g., gentle edges) are more stable.
Conventional learned image compression methods employ a standard MSE based R-D loss that treats all high-frequency residuals uniformly, lacking an explicit mechanism for content-adaptive bit allocation. This uniform weighting may cause the network to under-allocate bits to perceptually critical yet uncertain regions, sacrificing local fidelity to minimize global distortion.
}

To mitigate the impact of uncertainty from diffusion predictions on residual compression, we propose an uncertainty-guided residual compression module. The predicted high frequency $\Tilde{x}_{h}$ is sampled from the trained diffusion model, simultaneously its Bayesian uncertainty is estimated via Last-Layer Laplace Approximation (LLLA).
We then design a novel uncertainty-weighted R-D loss that dynamically prioritizes high uncertainty residuals, allocating more bits to them.

\noindent\textbf{Uncertainty estimation.} 
We follow BayesDiff approach~\cite{kou2023bayesdiff} to generate prediction high frequency and estimate its bayesian uncertainty simultaneously. Starting from the intial noisy image, the LLLA~\cite{daxberger2021laplace} is utilized for efficient Bayesian inference of pre-trained noise prediction models. Then the variance of sampled image is inferred along the reverse diffusion process until the final uncertainty map is estimated. 

The noise prediction model is trained to minimize Equation (\ref{eq:diffusion_loss}) under a weight decay regularizer, which corresponds to the Gaussian prior on the neural network parameters.
\begin{equation}
    p\left(\epsilon_{t} \mid x_{t}, t, \mathcal{D}\right) \approx \mathcal{N}\left({\epsilon}_{\theta}\left(\mathbf{x}_{t}, t,\bar{x}_{h}\right), \operatorname{diag}\left(\gamma_{\theta}^{2}\left(\boldsymbol{x}_{t}, t\right)\right)\right)
\end{equation}
where $\theta$ denotes parameters of the pre-trained noise prediction model.
 We keep only the diagonal elements in the Gaussian covariance $\gamma_{\theta}^{2}\left(\boldsymbol{x}_{t}, t\right)$, because they characterizes the Bayesian uncertainty over model parameters.
LLLA further improves the efficiency of bayesian uncertainty estimation by concerning only the parameters of the last layer of the neural network.

Next, we elaborate on integrating the uncertainty obtained above into the reverse diffusion process.
The sampling phase can commence with {$X_T \sim \mathcal{N}(0,\mathrm{\mathbf{I}})$} using the predicted noise  {$\epsilon_{\theta}\left(x_{t}, t, \bar{x}_{h}\right)$}, as follows:
\begin{equation}
    \mathbf{x}_{t-1}=\frac{1}{\sqrt{\alpha_{t}}}\left(\mathbf{x}_{t}-\frac{\beta_{t}}{\sqrt{1-\bar{\alpha}_{t}}}  {\epsilon}_{\theta}\left(\mathbf{x}_{t}, t,\bar{x}_{h}\right)\right)+\sigma_{t}  {z}
\label{Eq:get_xt-1}
\end{equation}

where {$ z \sim \mathcal{N}  (0,\mathrm{\mathbf{I} })$}, {$\alpha _t = 1 - \beta _t$}, and {$\bar\alpha _t =  {\textstyle \prod_{i=1}^{t}} \alpha _i$}. 

To estimate the pixel-wise uncertainty of $x_{t-1}$, we apply variance estimation to both sides of Eq.(\ref{Eq:get_xt-1}), giving rise to

\begin{equation}
\begin{split}
\operatorname{Var}\left(\mathbf{x}_{t-1}\right) = 
&\ \frac{1}{\alpha_{t}} \operatorname{Var}\left(\mathbf{x}_{t}\right) 
- 2 \frac{\beta_t}{\alpha_{t} \sqrt{1-\bar{\alpha}_{t}}} \operatorname{Cov}\left(\boldsymbol{x}_{t}, \boldsymbol{\epsilon}_{t}\right) \\
&\ + \frac{\beta_t^{2}}{\alpha_{t}\left(1-\bar{\alpha}_{t}\right)} \operatorname{Var}\left(\boldsymbol{\epsilon}_{t}\right) 
+ \sigma_{t}^2
\label{Eq:varx}
\end{split}
\end{equation}
where $Cov(x_{t},\epsilon _{t})$ denotes the covariance between $x_t$ and $\epsilon_t$. 

With this, we can iterate over it to estimate the pixel-wise uncertainty of the final $x_0$, i.e., Var($x_0$). 
Recalling that $Var(\epsilon_t)=\gamma _{\theta }^{2}(x_t,t) $ which has been estimated by LLLA, the main challenges then boils down to estimating $Cov(x_{t},\epsilon _{t})$.

By the law of expectation $E(E(X|Y )) = E(X)$, there is
\small
\begin{equation}
\begin{split}
\operatorname{Cov}\left(\boldsymbol{x}_{t}, \boldsymbol{\epsilon}_{t}\right) & =E\left(\boldsymbol{x}_{t} \odot \boldsymbol{\epsilon}_{t}\right)-E\boldsymbol{x}_{t} \odot E \boldsymbol{\epsilon}_{t} \\
& =E_{\boldsymbol{x}_{t}}\left(\boldsymbol{x}_{t} \odot \epsilon_{\theta}\left(\boldsymbol{x}_{t}, t\right)\right)-E \boldsymbol{x}_{t} \odot E_{\boldsymbol{x}_{t}}\left(\epsilon_{\theta}\left(\boldsymbol{x}_{t}, t\right)\right)    
\end{split}
\label{Eq:getCov}
\end{equation}

where $\odot $ denotes the element-wise multiplication.
It is straightforward to estimate $E\boldsymbol{x}_{t}$ via a similar iteration rule
\begin{equation}
   E({x_{t - 1}}) = \frac{1}{{\sqrt {{a_t}} }}E({x_t}) - \frac{{\beta_{t}}}{{\sqrt {{{a}_t}(1 - {\bar{a}_t})} }}E({\varepsilon _t})
\label{Eq:getExt-1}
\end{equation}
Given these, we can reasonably assume $x_t$ follows $N(E({x_t}),Var({x_t}))$ , and then $Cov(x_t,\epsilon_t)$ can be approximated with Monte Carlo (MC) estimation:
\begin{equation}
    Cov({x_t},{\varepsilon _t}) \approx \frac{1}{S}\sum\limits_{i = 1}^S {({x_{t,i}},t) - E{x_t}}  \odot \frac{1}{S}\sum\limits_{i = 1}^S {{\varepsilon _\theta }({x_{t,i}},t)}
    \label{eq:cov}
\end{equation}

Substituting Eq.(\ref{eq:cov}) into Eq.(\ref{Eq:varx}), the uncertainty of sampling $x_{t-1} $ is estimated iteratively until the pixel-wise uncertainty $Var(x_0)$. Alg.\ref{alg:Wavelet}. demonstrates the procedure of applying the developed uncertainty iteration principle to the diffusion model. 


\begin{algorithm}[t]
\caption{Wavelet Diffusion Inference with Uncertainty Estimation}
\label{alg:Wavelet}
\begin{algorithmic}[1] 
\REQUIRE ~~\\ 
    synthetic high frequency image {$\bar{x}_{h}$}; pre-trained noise prediction network {$\epsilon_{\theta}$};\\
\ENSURE ~~\\ 
    Predicted high frequency \textit{$\Tilde{x}_{h}$}; uncertainty map Var(\textit{$\Tilde{x}_{h}$});
    \STATE {$ x_T \sim \mathcal{N}  (0,\mathrm{\mathbf{I} })$} 
    \STATE {Construct the variance prediction function $\gamma_{\theta}^{2}$ via LLLA}
    \STATE {$E\left(\boldsymbol{x}_{T}\right) \leftarrow \boldsymbol{x}_{T}, \operatorname{Var}\left(\boldsymbol{x}_{T}\right) \leftarrow \mathbf{0}, \operatorname{Cov}\left(\boldsymbol{x}_{T}, \boldsymbol{\epsilon}_{T}\right) \leftarrow \mathbf{0}$}
    \FOR{$t=T:1$ }
        \STATE {
            Sample  $\boldsymbol{\epsilon}_{t} \sim \mathcal{N}\left(\epsilon_{\theta}\left(\boldsymbol{x}_{t}, t\right), \operatorname{diag}\left(\gamma_{\theta}^{2}\left(\boldsymbol{x}_{t}, t\right)\right)\right)$
        }
        \STATE {
            {$ z \sim \mathcal{N}  (0,\mathrm{\mathbf{I} })$} if \textit{t} {$> 1$}, else {$z = 0$} 
        }
        \STATE {
            Obtain $\boldsymbol{x}_{t-1}$ via Eq.(\ref{Eq:get_xt-1})
        }
        \STATE {
            Estimate $E\left(\boldsymbol{x}_{t-1}\right)$ and $\operatorname{Var}\left(\boldsymbol{x}_{t-1}\right)$ via Eqs.(\ref{Eq:getExt-1}) and(\ref{Eq:varx})
        }
        \STATE {
            Sample  $\boldsymbol{x}_{t-1, i}\sim\mathcal{N}\left(E\left(\boldsymbol{x}_{t-1}\right), \operatorname{Var}\left(\boldsymbol{x}_{t-1}\right)\right), i = 1, \ldots, S $
        }
        \STATE {
            Estimate  $\operatorname{Cov}\left(\boldsymbol{x}_{t-1}, \boldsymbol{\epsilon}_{t-1}\right)$ via Eq. (\ref{Eq:getCov}).
        }
    \ENDFOR
\RETURN {$\Tilde{x}_{h} = x_0$}, {Var$(\textit{$\Tilde{x}_{h}$})$=Var($x_0$)}
\end{algorithmic}
\end{algorithm}

\noindent\textbf{Uncertainty weighted R-D Loss.} 
\textcolor{blue}{The aleatoric uncertainty map
$\delta$ estimated from the diffusion prediction quantifies the reliability of the high-frequency prediction: higher 
$\delta$ indicates lower confidence, implying that more bits should be allocated to ensure fidelity.
We will introduce an uncertainty-related weight to the standard MSE distortion loss to prioritize residuals with high uncertainty.
}

Given an arbitrary image $x$, optimizing the VAE based image compression model for R–D performance has been proven to be equivalent to minimization of the KL divergence as follows~\cite{balle2018variational},
\begin{equation}
\begin{aligned}
L_{RD} &\propto E_{x\sim p_{x}  } D_{KL} [q||p_{\tilde{y}|x } ]
\\&=\mathbb{E} _{x\sim p_{x}} \mathbb{E}_{\tilde{y} \sim q} \left [ \underbrace{-logp_{x|\tilde{y}}(x|\tilde{y})}_{\text{weighted distortion}}\underbrace{-logp_{\tilde {y}} (\tilde {y})}_{\text{rate}}   \right ]
\end{aligned}
\label{eq:rd}
\end{equation}
where $\tilde{y}$ is an approximation to the quantized latent representations $\hat{y}$ with an additive i.i.d. uniform noise to enable end-to-end training. 

Specifically, minimizing the first term in KL divergence is equivalent to minimizing the distortion between original $x$ and reconstructed $\tilde{x}$ measured by squared difference when the likelihood $ p_{\boldsymbol{x} \mid \tilde{\boldsymbol{y}}}\left(\boldsymbol{x} \mid \tilde{\boldsymbol{y}}\right)\sim \mathcal{N} (x|\tilde{x},(2\lambda )^{-1}I )$. The second term in Eq.(\ref{eq:rd}) denotes the cross entropy that reflects the cost of encoding $\tilde{y}$ i.e., bitrate $R$.
The R-D loss is:
\begin{equation}
L_{RD} = R+\lambda \left \| x-\tilde {x} \right \| _{2}
\label{eq:Lrd}
\end{equation}
where $\lambda$ is a hyper-parameter used to balance the overall rate-distortion, i.e.,larger $\lambda$ for larger rate and better reconstruction quality and vice versa.

\begin{figure*}[t]  
\centering  
\includegraphics[width=1.0\linewidth]{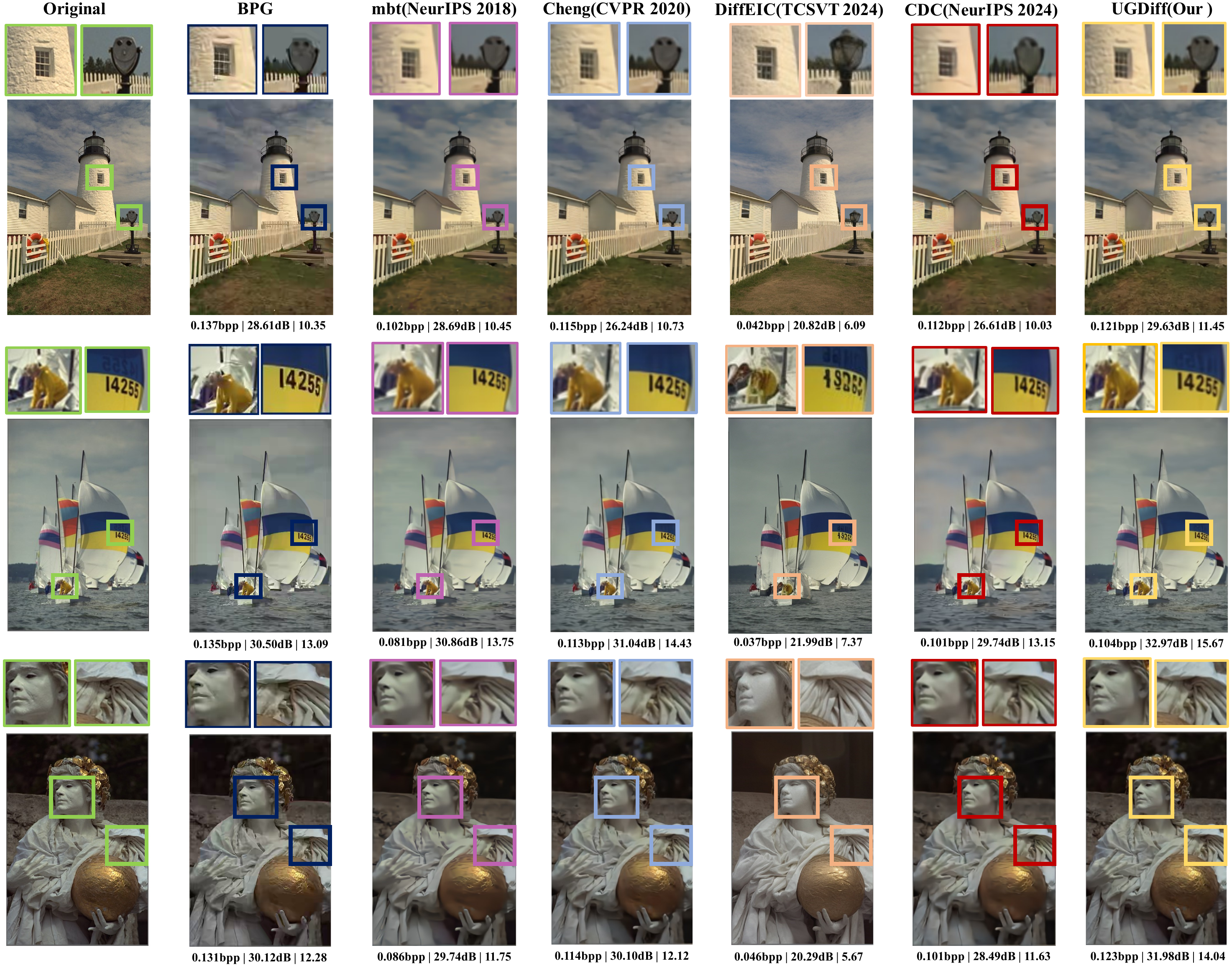}  
\vspace{-6mm}
\caption{Visualization of the reconstructed images from Kodak dataset. The metrics are [bpp↓/PNSR↑/MS-SSIM↑].} 
\label{fig:kodak19}  
\vspace{-5mm}
\end{figure*}

Standard MSE based R-D loss that treats all high-frequency residuals uniformly, lacking an explicit mechanism for content-adaptive bit allocation. That may lead the network
to sacrifice perceptually critical regions to minimize overall
distortion.
To address this issue, we reconsider the weighted R-D loss with aleatoric uncertainty. Let $x_r$ represent the residuals to be compressed, $f(\cdot)$ represents the variational inference in residual compression module, $\delta$ denotes the aleatoric uncertainty estimated in the above subsection. This way, the compression model can be formulated as:
\begin{equation}
    {x}_r = f(\Tilde{y}) + \epsilon\delta
\end{equation}
where $\epsilon$ represents the Gaussian distribution with zero-mean and unit-variance, assumed for characterizing the likelihood function by:
\begin{equation}
    p\left({x}_{r} \mid \Tilde{y}, \delta\right)  = \frac{1}{\sqrt{2 \pi \delta}} \exp \left(-\frac{\left\|{x}_{r}-f(\Tilde{y})\right\|_{2}}{2 \delta}\right)
\label{eq:distribution}
\end{equation}
Then a negative log likelihood then works out to be the uncertainty weighted distortion term between ${x}_{r}$ and $f(\Tilde{y})$,
\begin{equation}
-\log\left( p\left({x}_{r} \mid \Tilde{y}, \delta\right) \right) \propto \frac{\left\|{x}_{r}-f(\Tilde{y})\right\|_{2}}{2 \delta}
\end{equation}
A naive formulation using the negative log-likelihood of the residual $x_r$ yields a distortion term proportional to $\frac{\left\|x_{r}-f(\tilde{y})\right\|_{2}}{2 \delta}$.
We observe that the uncertainty-derived weight $(2\delta)^{-1}$ in the  distortion term act as penalties on regions with high uncertainty.
However,this formulation is counter-intuitive from a compression perspective:regions with higher uncertainty (larger $\delta $), should be prioritized with increased bit allocation rather than penalized.

To align the optimization objective with our design principle, we seek a monotonically increasing function to prioritize pixels with large uncertainty rather than penalize them using  $(2\delta)^{-1} $.

Differential entropy measures the information content of a continuous random variable, which reflects the cost of coding. The differential entropy of the random variable $X$ is computed as follows 
\begin{equation}
    H(X) = -\int p(X)\log(p(X))dX
    \label{eq:entropy}
\end{equation}
 We substitute the probability distribution in Eq.(\ref{eq:distribution}) into Eq.(\ref{eq:entropy}) to obtain the differential entropy $H(x_{r})$ of $x_{r}$, 
\begin{equation}
    H(x_{r})=\log (\delta \sqrt{2 \pi})
    \label{eq:H_gaussian}
\end{equation}
Eq.(\ref{eq:H_gaussian}) demonstrates the increase trend of differential entropy with the variance $\delta$. 

Motivated by this equation, we design a new adaptive weighted loss named uncertainty-weighted rate-distortion loss ($L_{URD}$), in which the weight $log(\delta)$ is used to prioritize pixels with large uncertainty in the R-D loss function. 
Combining the hyper-parameter $\lambda$ to balance the overall trade-off between the rate and distortion, the uncertainty weighted R-D loss function is reformulated as:
\begin{equation}
    L_{URD} = R+ (\lambda + log(\delta)) \cdot \left\|{x}_{r}- \tilde{{x}_{r}}\right\|_{2}
\label{eq:residual_loss}
\end{equation}

where $\lambda$  globally controls the overall rate-distortion trade-off, determining the total bitrate budget, whereas  estimated uncertainty $log(\delta)$ serves as the content adaptive weight to prioritize pixels with large uncertainty and allocate more bits to them during compression.  
Our proposed $L_{URD}$ enables a more rational allocation of the fixed bitrate budget, allocating more bits to perceptually critical regions where diffusion prediction is less reliable, thereby achieving a superior rate-distortion-perception balance.

\subsection{Training Strategy}
As the analysis above, the whole training process of UGDiff contains four steps.
\textcolor{blue}{Firstly, we train a learned image compression network~\cite{minnen2020channel} for our low frequency codec.}

The loss function is: 
\begin{equation}
\begin{aligned}
    {L_{low}} & = R+\lambda_l \cdot D \\
\end{aligned}
\label{eq:codec_loss}
\end{equation}
where {$\lambda_l$} controls the trade-off between rate and distortion. \textit{R} represents the bit rate of latent \textit{$\hat{y}$} and side information \textit{$\hat{z}$}, and $D$ 
is the MSE distortion term.

\textcolor{blue}{The second step is to train the low-to-high frequency translator
 $G_{\psi}$, 
optimized by minimizing MSE between the output and the original high frequency, which learns to map compressed low-frequency to  synthetic high-frequency. 
\textcolor{blue}{During this process, the parameters of the low-frequency codec are frozen.}
}
\begin{equation}
\ {L_{trans}} = ||G_{\psi}\left(\hat{x}_{l}\right)-{{x}_{h}}\|_2
\label{eq:domain_loss}
\end{equation}


\begin{figure*}[t!]  
\centering  
\includegraphics[width=1.0\linewidth]{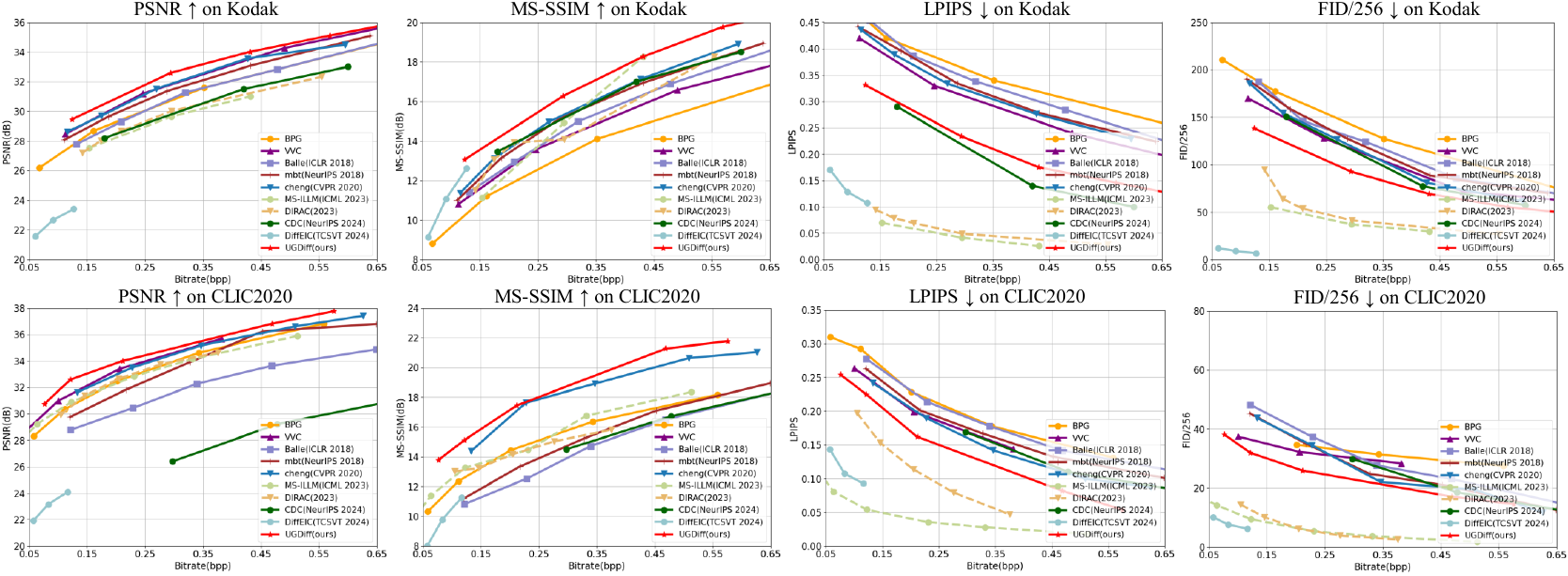} 
\vspace{-5mm}
\caption{Quantitative comparisons with SOTA methods in terms of perceptual quality (PSNR$\uparrow$ / MS-SSIM$\uparrow$ / LPIPS$\downarrow$ / FID/256$\downarrow$) on Kodak~\cite{franzen1999kodak} and CLIC2020~\cite{clic2020} datasets. }  
\label{fig:result_kodak_clic2020}  
\vspace{-5mm}
\end{figure*}

The third step is to train the conditional diffusion model by optimizing Eq.(\ref{eq:diffusion_loss}). During this process, we freeze the low-frequency encoder-decoder network and the low-to-high frequency translator
module $G_{\psi}$, ensuring that only the diffusion model is optimized. The design of the noise prediction network in the diffusion model follows a similar U-Net architecture used in DDPM~\cite{ho2020denoising}, enhanced with residual convolutional blocks~\cite{He_2016_CVPR} and self-attention mechanisms.



The final step is to train the uncertainty-guided residual compression model. 
The residual compression model follows the same structure as the low-frequency compression network, but is specifically trained to focus on minimizing the residual loss, incorporating uncertainty weighting for enhanced performance.
During this training process, we freeze the low-frequency compression network, the low-to-high frequency translator and the conditional diffusion model; only the residual compression network is optimized to minimize the uncertainty-weighted R-D loss in Eq.(\ref{eq:residual_loss}). 

\section{Experimental Results}
\subsection{Implement Details}

\noindent\textbf{Training.}
\textcolor{blue}{We train all components of our framework on the OpenImages dataset~\cite{openimages}, from which we randomly sample 300k images per epoch, resized to 256×256. Specifically, the frequency translator is trained with MSE loss between the synthetic and original high frequency using Adam optimizer ~\cite{KingBa15} for 200 epoches with a batch size of 32. The learnig rate was set to {$1\times 10^{-4}$}. The wavelet diffusion model is trained on the same 256×256 high-frequency subbands, At each iteration, a time step $t$ is sampled uniformly, and Gaussian noise is added to the clean high-frequency. The network is optimized using Adam optimizer for 200 epochs  with a batchsize of 32. The learning rate was set to {$5\times 10^{-4}$}.The sampling step was set to $T=10$. 
Low-frequency and residual codec are trained separately for 1.8M steps with a batch size of 16 using a multi-stage learning rate:{$1\times 10^{-4}$}(first 120k), {$3\times 10^{-5}$} (next 30k), and {$1\times 10^{-5}$} (last 30k). We set $\lambda_l,\lambda\in\left \{0.01,0.05,0.1,0.2,0.3\right\} $for MSE-based R-D optimization.
All models use base feature channels of encoder/decoder backbone $N=192$ and compressed latent channels $M=320$  following~\cite{minnen2020channel}.}
Experiments are implemented in PyTorch~\cite{paszke2019pytorch} and CompressAI~\cite{begaint2020compressai} on an NVIDIA RTX 4090 GPU.

\begin{table}[t] 
\caption{\textcolor{blue}{BD-rate ($\downarrow$) savings with VVC as the anchor on Kodak and CLIC2020 datasets. BD-rate is computed based on PSNR. Negative BD-rate values indicate better performance.}}
\centering
\small
\scalebox{0.8}{
    \begin{tabular}{lcc}
        \toprule
        Method & BD-rate on Kodak $\downarrow$ & BD-rate on CLIC2020 $\downarrow$ \\
        \midrule
        VVC(~\cite{vvc})  & 0\% & 0\% \\
        BPG(~\cite{bpg}) & 11.43\% & 6.73\% \\
        Cheng'2020(~\cite{cheng2020learned}) & 2.82\% & 1.37\% \\
        MS-ILLM'2023(~\cite{2023-ICML-MS-ILLM}) & 19.28\% & 6.37\% \\
        DIRAC'2023(~\cite{liu2023learned}) & 17.35\% & 6.42\% \\
        CDC'2024(~\cite{yang2024lossy}) & 15.64\% & 19.59\% \\
        \textbf{Ours UGDiff} & \textbf{-8.02\%} & \textbf{-18.4}\% \\
        \bottomrule
    \end{tabular}
}

\label{table:bdrate}
\vspace{-2mm}
\end{table}

\noindent\textbf{Evaluation.}
\textcolor{blue}{The evaluations are conducted on the Kodak dataset~\cite{franzen1999kodak} and CLIC2020 test dataset~\cite{clic2020}. 
The Kodak dataset comprises 24 images with a resolution of 768x512 pixels. For CLIC2020, images are resized to a minimum dimension of 768px and center-cropped to 768×768 for evaluation.
More experimental results on Tecknick dataset~\cite{tecnick} can be found in supplementary materials.
We reach different ranges of bitrates by compressing images with different models trained using different $\lambda$ and $\lambda_l$. 
PSNR and MS-SSIM~\cite{1292216} metrics are computed to evaluate distortion loss. 
In addition, we also compute the Learned Perceptual Image Patch Similarity (LPIPS) metric~\cite{8578166} and Frechet Inception Distance (FID/256)~\cite{2017-FID}.}

\noindent\textbf{Baselines.}
\textcolor{blue}{To show the effectiveness of our UGDiff, we compare its R-D performance with SOTA image compression methods. Traditional compression standards include BPG~\cite{bpg}, and VVC~\cite{vvc}. Learned compression methods include context-free hyperprior model (Balle ICLR2018)~\cite{balle2018variational}, auto-regressive hyperprior models (mbt NeurIPS2018)~\cite{minnen2018joint}, entropy models with Gaussian Mixture Models and simplified attention (Cheng CVPR2020)~\cite{cheng2020learned} and Mean-Scale-ILLM (MS-ILLM ICML2023)~\cite{2023-ICML-MS-ILLM}. Diffusion model-based compression methods include CDC NeurIPS2024 ($\rho=0$)~\cite{yang2024lossy}, DIRAC(single sampling step is adopted to achieve minimal distortion)~\cite{ghouse2023residual} and DiffEIC TCSVT2024 ~\cite{2024Towards}.
Comparisons with some more SOTA image compression methods can be found in Supplementary Materials.
}

\subsection{Comparison with the SOTA Methods}
\noindent\textbf{Rate-Distortion Performance.}  
\textcolor{blue}{Fig.~\ref{fig:result_kodak_clic2020} 
presents a comprehensive quantitative comparison of UGDiff against SOTA image compression methods on the Kodak and CLIC2020 benchmarks, evaluating performance across a wide range of bitrates in terms of PSNR, MS-SSIM, LPIPS, and FID/256.
The results show that UGDiff consistently outperforms SOTA methods in terms of PSNR and MS-SSIM metrics and also achieves competitive perceptual performance in terms of LPIPS and FID/256 metrics, covering traditional codecs, learned image compression approaches, and existing diffusion-based methods. For instance, UGDiff achieves an average PSNR gain of 0.8 dB over Cheng's method at bitrate 0.3bpp. This confirms the effectiveness of our wavelet diffusion prediction + uncertainty-guided residual compression paradigm in preserving pixel-level fidelity. While perceptual methods, such as MS-ILLM and diffusion based DiffEIC, achieve strong perceptual quality, they typically exhibit inferior PSNR and MS-SSIM performance. In contrast, UGDiff achieves a more favorable balance between perceptual quality and reconstruction fidelity across all evaluated bitrates. Experimental results compared with more SOTA baselines on more dataset can be found in supplementary materials.
}


\noindent\textbf{BD-rate Analysis.}
\textcolor{blue}{To compare R-D performance quantitatively, we use the BD-rate metric~\cite{bdrate} to calculate average bitrate savings at the same PSNR quality. Using VVC intra~\cite{vvc} (version 12.1) as the anchor, BD-rates are shown in Table \ref{table:bdrate}. UGDiff achieves the highest BD-rate savings, with 8.02\% and 18.4\% savings on the Kodak and CLIC2020 datasets compared to VVC respectively.
UGDiff achieves BD-rate savings of 23.66\% and 31.13\% compared to the CDC diffusion-based approach, showing significant improvement in distortion metrics. These results highlight UGDiff’s superior pixel-level fidelity over SOTA image compression methods.}

\noindent\textbf{Subjective Quality Comparison.} 
We also perform subjective quality evaluations on the Kodak dataset. Fig.\ref{fig:kodak19} shows visual comparisons between original images and those reconstructed by various compression methods. 
BPG introduces noticeable ringing and quantization artifacts around edges and textured regions (second column, Fig.~\ref{fig:kodak19}), which degrade fine structural details. Learned methods, like mbt'2018 and Cheng'2020, suffer from over-smoothing, losing textural fidelity, with details like the smile on the signboard or numerals on the sail becoming obscured. Diffusion methods, such as DiffEIC'2024, achieve good perceptual quality but suffer from semantic inconsistencies, like misinterpreting a signboard as a street lamp. 
UGDiff, however, retains more high-frequency details and superior visual quality, preserving fine details such as smiles, numerals, and facial features.

\noindent\textbf{Complexity Analysis}. 
We evaluate the complexity by comparing the inference time of different compression methods on the Kodak dataset (768×512). 
Encoding and decoding times are calculated at similar R-D points to assess model complexity. For fairness, all models are implemented on the same GPU using their public codes. 
As shown in Table \ref{table:time}, Balle'2018 exhibits the lowest complexity among the learned image codecs.
The CDC diffusion model suffers from slow decoding due to its iterative denoising process (500 sampling steps), taking about 55s to decode an image. In contrast, UGDiff, using a wavelet diffusion model applied only to sparse high-frequency components, is at least 40× faster, requiring only 10 sampling steps. This reduces decoding time from 55s to 1.47s, with a higher PSNR than CDC.

\begin{table}[t]
\centering
\caption{Comparison of the averaged encoding and decoding time on Kodak dataset.}
\footnotesize
\scalebox{0.9}{
    \begin{tabular}{lcccc}
\toprule
Methods & Enc (s) & Dec (s) & PSNR $\uparrow$ & bpp $\downarrow$ \\
\midrule
BPG(~\cite{bpg}) & 0.66 & 0.17 & 34.85 & 0.68 \\
VVC(~\cite{vvc}) & 102.5 & 0.12 & 34.26 & 0.50 \\
Cheng'2020(~\cite{cheng2020learned}) & 5.40 & 9.25 & 34.94 & 0.595 \\
Balle'2018(~\cite{balle2018variational}) & 0.49 & 0.62 & 34.72 & 0.668 \\
mbt'2018(~\cite{minnen2018joint}) & 7.82 & 10.41 & 35.09 & 0.638 \\
CDC'2024(~\cite{yang2024lossy}) & 0.53 & 55.47 & 33.01 & 0.598 \\
Our UGDiff & 2.01 & 1.47 & \textbf{35.46} & 0.635 \\
\bottomrule
    \end{tabular}
}
\label{table:time}
\vspace{-7mm}
\end{table}

\subsection{Ablation Studies}
\textcolor{blue}{We conduct the ablation studies to further analyze our proposed UGDiff.
Firstly, We compare 5 variants in Table \ref{table:bdrate-ablation} to evaluate the impact of different components on image compression performance in terms of BD-rate through incrementally including each specific component. 
The specific components encompasses the low frequency codec, Wavelet Diffusion, Frequency Translator, Residual Compression and Uncertainty Guidance. The baseline low-frequency codec, which reconstructs the image only from the low frequency, is set as the anchor to compute BD-rate.}

\begin{table}[t]
\caption{\textcolor{blue}{Ablation study of different components on the Kodak and CLIC2020 datasets. BD-rate is computed based on PSNR. Each row incrementally adds one module to the previous configuration. When the residual compression module is absent, high frequency is directly reconstructed from the  diffusion module.}}
\centering
\small
\scalebox{0.75}{
    \begin{tabular}{lcc}
        \toprule
        Component & BD-rate on Kodak $\downarrow$& BD-rate on CLIC2020 $\downarrow$\\
        \midrule
        Low Frequency Codec(Baseline) & {0\%} & {0\%}\\
       +Wavelet Diffusion(direct reconstruction) & {-19.53\%} & {-12.79\%}\\
        +Frequency Translator & {-30.38\%} & {-20.36\%}\\
        +Residual Compression & {-80.41\%} & {-64.27\%}\\
        +Uncertainty Guidance & {-85.23\%} & {-71.82\%}\\
        \bottomrule
    \end{tabular}
}
\label{table:bdrate-ablation}
\vspace{-2mm}
\end{table}

\begin{table}[t] 
\centering
    \caption{Ablation studies of various settings on the condition and sampling step on the Kodak dataset.}
    \scalebox{0.8}{
    {$
        \begin{array}{c|c|ccccc}
            \hline 
            \text { Settings } & \text { Step }&\text { PSNR } \uparrow&\text{ MS-SSIM } \uparrow &\text { LPIPS } \downarrow&\text { Times (s) } \downarrow & \text{bpp} \downarrow \\
            \hline 
            &T=1 & 24.67 & 13.92 & 0.468 & 0.64 & 0.646\\
            & T=5 & 34.86 & 19.54 & 0.21 & 0.95 & 0.633\\
            & T=10 & \textbf{35.46} & \textbf{20.26} & \textbf{0.18} & \textbf{1.47} & \textbf{0.635}\\
         \text{UGDiff}(\Bar{x}_{h})   & T=20 & 35.49 & 20.21 & 0.18 & 2.65 & 0.632\\
            &T=30 & 35.37 & 19.89 & 0.19 & 3.48 & 0.633\\
            &T=50 & 35.42 & 19.97 & 0.18 & 6.04 & 0.635\\
            \hline
            &T=1 & 25.14 & 12.95 & 0.453 & 0.69 & 0.644\\
            &T=5 & 33.85 & 18.99 & 0.22 & 0.97 & 0.642\\
            & T=10 & \textbf{35.26} & \textbf{19.01} & \textbf{0.21} & \textbf{1.45} & \textbf{0.641}\\
        \text{UGDiff}(\hat{x}_{l})    &T=20 & 34.91 & 18.97 & 0.22 & 2.75 & 0.642\\
            &T=30 & 35.28 & 19.04 & 0.21 & 3.79 & 0.640\\
            &T=50 & 35.19 & 18.99 & 0.21 & 6.18 & 0.643\\
            \hline
            &T=1 & 12.87 & 2.12 & 0.91 & 0.18 & 0.723\\
            &T=5 & 15.69 & 6.89 & 0.28 & 0.33 & 0.723\\
            & T=10 & 27.56 & 16.96 & 0.16 & 0.77 & 0.723\\
            \text{CDC 2024} &T=65 & 33.57 & 20.06 & 0.14 & 7.05 & 0.723\\
            &T=100 & 34.06 & 20.17 & 0.14 & 10.89 & 0.723\\
            &T=500 & 34.46 & 20.21 & 0.13 & 55.47 & 0.723\\
            \hline
        \end{array}
    $}
    }
\label{table:sampling}
\vspace{-5mm}
\end{table}


\noindent\textbf{Effect of Wavelet Diffusion.} 
As shown in Table \ref{table:bdrate-ablation},the introduction of the wavelet diffusion module for direct high-frequency reconstruction yields a substantial improvement over the baseline (low-frequency codec), achieving BD-rate savings of 19.53\% on Kodak and 12.79\% on CLIC2020. This performance gain underscores the generative capabilities of  diffusion model in synthesizing  high-frequency details.

\noindent\textbf{Effect of Frequency Translator.} 
\textcolor{blue}{By comparing the second and third rows in Table \ref{table:bdrate-ablation}, it can be observed that additional BD-rate savings of 10.85\% on the Kodak dataset and 7.57\% on the CLIC2020 dataset are achieved when using synthetic high-frequency produced by the frequency translator instead of reconstructed low frequency as the condition of diffusion model. It indicates that the synthetic high-frequency produced by frequency translator provides more informative guidance for condition diffusion than low-frequency.}

\noindent\textbf{Effect of Residual Compression.} 
\textcolor{blue}{By comparing the third and forth row in Table \ref{table:bdrate-ablation}, it can be observed that additional BD-rate savings of 50.03\% on  Kodak dataset and 43.91\% on CLIC2020 dataset are achieved when wavelet diffusion is applied for prediction, and the prediction residual is transmitted to the decoder. The core mechanism behind this performance leap is  the correction of synthetic errors of diffusion models through transmitting residuals. This"prediction-then-residual-compression" strategy directly addresses the inherent low-fidelity issue of vanilla diffusion models.}

\noindent\textbf{Effect of Uncertainty Guidance.} By comparing the forth and final row in the Table.\ref{table:bdrate-ablation}, it can be observed that BD-rate savings of 4.82\% on the Kodak dataset and 7.55\% on the CLIC2020 dataset are achieved when the uncertainty of diffusion model is introduced in the  R-D loss of residual compression. It indicates that our dynamic, content-aware uncertainty-weighted R-D loss enables a more rational rate-distortion trade-off than a uniform R-D loss, ultimately leading to superior compression efficiency.


\begin{figure}[t]  
\centering  
\includegraphics[width=1\linewidth]{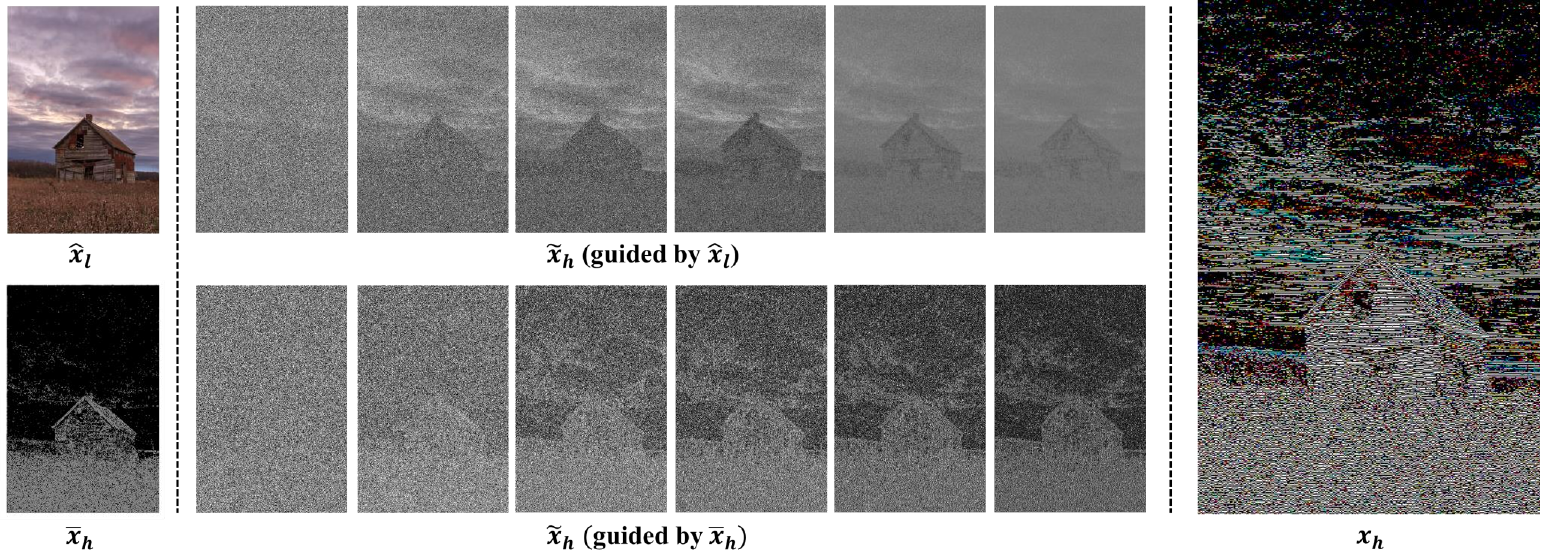}  
\vspace{-5mm}
\caption{Visualization results of reverse diffusion process under different conditions. The left part displays the different conditions \textit{$\hat{x}_{l}$} and \textit{$\bar{x}_{h}$}, the middle part illustrates the reverse diffusion process under different conditions, and the right part exhibits the original high frequency information {$x_h$}. From the 10 sampling steps, we selected the generated results \textit{$\Tilde{x}_{h}$} at t = {10, 8, 6, 4, 2, 0} for visualization.}  
\label{x_l_guide}  
\vspace{-5mm}
\end{figure} 

\begin{figure}[htbp]
\centering
\includegraphics[width=1\linewidth]{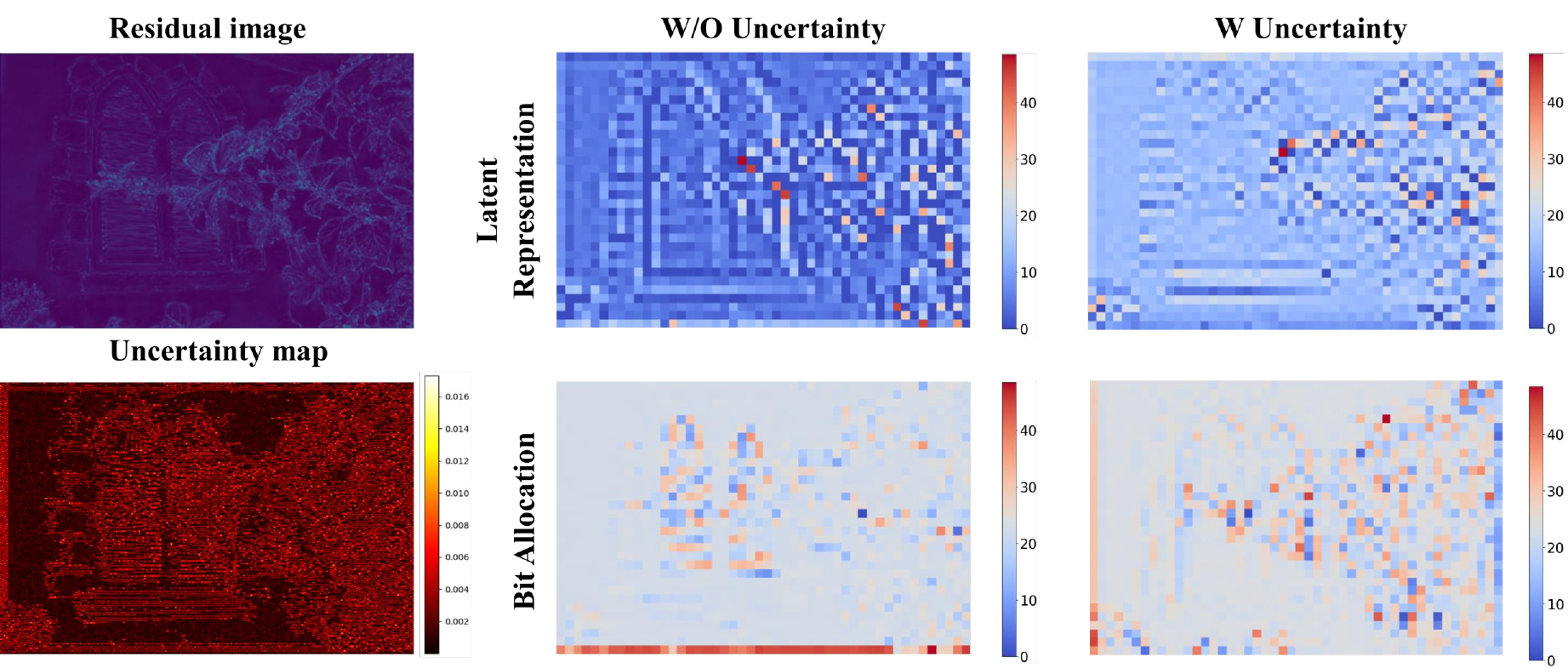}
\vspace{-3mm}
\caption{Visualization of the estimated uncertainty map, the latent representation, and the bit allocation of the model with and without R-D optimization guided by uncertainty.}
\label{fig:bit-ablation}
\vspace{-4mm}
\end{figure}

\noindent\textbf{Effect of Sampling steps.} 
Table.\ref{table:sampling} compares the overall R-D performance of our proposed UGDiff under different conditions and sampling steps.
The results reveal that UGDiff’s performance saturates at T=10, regardless of whether the diffusion is conditioned on the reconstructed low-frequency component $\hat{{x}}_{l}$
or the synthetic high-frequency $\bar{x}_{h}$ .
By contrast, CDC 2024 model requires T=500
steps to achieve its optimal perceptual quality. 
This dramatic difference in convergence behavior highlights the efficiency of our wavelet-domain approach: by operating solely on the sparse high-frequency subbands which contain more sparse information compared to the image domain.

\subsection{Visualizations }

\noindent\textbf{Wavelet diffusion condition.}
Fig.\ref{x_l_guide} provides a direct visual comparison of the reverse diffusion process under two distinct conditions: the reconstructed low-frequency component \textit{$\hat{x}_{l}$} and the synthetic high-frequency \textit{$\bar{x}_{h}$}.From the first row of the figure, it is evident that conditioning the diffusion model on \textit{$\hat{x}_{l}$} , leads to producing an image that resembles the smooth, low-resolution structure of the input, consequently manifesting a loss of certain detailed textures. That would make the prediction residual quite large and affect the efficacy of residual compression. 
The second row shows that the synthetic high-frequency components generated by the frequency translator contain more high-frequency details. 
The visualization of the reverse diffusion process indicates that images generated through reverse diffusion conditioned on synthetic high-frequency components exhibit a closer resemblance to the original high-frequency components compared to those conditioned on low-frequency components.

\noindent\textbf{Uncertainty-weighted rate–distortion loss.}
We visualize the effect of the proposed uncertainty-weighted rate–distortion loss on residual compression in Fig.~\ref{fig:bit-ablation}. The results show that the uncertainty map are well aligned with predictive residuals, indicating that the map captures the instability of the conditional diffusion model in high-frequency prediction. Using the standard rate–distortion loss (W/O Uncertainty), residuals are treated uniformly, resulting in evenly distributed bit allocation. In contrast, the uncertainty-weighted loss emphasizes large residuals highlighted by the uncertainty map, making their latent representations more prominent and enabling more efficient bit allocation with only a slight increase in bitrate.

\vspace{-1mm}
\section{Conclusion}


We propose UGDiff, an uncertainty-guided image compression method based on wavelet diffusion to strick a balance between high perceptual quality and low distortion. DWT is leveraged  to decouple the image into low-frequency and high-frequency, enabling a dedicated diffusion model for high frequency to predict fine details and a deterministic low-frequency codec to preserve global structure. Wavelet diffusion is utilized to predict rather than directly reconstruct the high frequency. The prediction residuals are then transmitted to the decoder. This diffusion prediction-then-residual compression effectively mitigates the low-fidelity issue of existing diffusion-based methods. We further introduce an uncertainty-weighted rate–distortion loss to achieve a rational R-D trade-off. Experimental results demonstrate that UGDiff outperforms SOTA learned compression methods in both R-D performance and visual quality. In the future, we will explore single-step diffusion models in the future to further improve efficiency.

\vspace{-2mm}
\section*{Acknowledgments}
This work was supported in part by the National Natural Science Foundation of China under Grant 62373293. 

\bibliographystyle{IEEEtran}
\bibliography{main}

\clearpage

\end{document}


\title{Supplementary Materials}

\markboth{Journal of \LaTeX\ Class Files,~Vol.~14, No.~8, 2024}%
{Shell \MakeLowercase{\textit{et al.}}: A Sample Article Using IEEEtran.cls for IEEE Journals}

\maketitle

\section{Comparative experiments on Tecnick}
To validate the generalization capabilities of our proposed method across different scenarios, we conducted a quantitative evaluation on the Tecnick dataset. The SOTA baseline include BPG~\cite{bpg}, VVC~\cite{vvc}, Balle ICLR2018 ~\cite{balle2018variational}, mbt NeurIPS2018~\cite{minnen2018joint},  Cheng CVPR2020~\cite{cheng2020learned}, CDC NeurIPS2024~\cite{yang2024lossy}. 
Figure \ref{fig:result_tecnick} illustrates the Rate-Distortion (RD) curve comparisons for the objective metric PSNR and the perceptual metric LPIPS.
As observed from the PSNR results (left), our proposed UGDiff (red curve) achieves state-of-the-art performance across the wide bitrate range , significantly outperforming traditional codecs (e.g., BPG, VVC), learning-based methods (e.g., Cheng2020, mbt) and diffusion based approach (CDC).
Notably, while the CDC method (green curve) yields lower values in the LPIPS perceptual metric (right) , this comes at the expense of a significant sacrifice in pixel-level fidelity,  its PSNR performance is substantially lower than all other competing methods. In contrast, UGDiff not only maintains a substantial lead in PSNR, but also achieves highly competitive results in LPIPS, ranking second only to CDC and outperforming other baselines. This demonstrates that UGDiff more effectively balances pixel fidelity with perceptual quality, exhibiting superior robustness across different data distributions.

\begin{figure}[h!]  
\centering  
\includegraphics[width=1.0\linewidth]{figs/tecnick_psnr_lpips.pdf} 
\vspace{-5mm}
\caption{Comparison experiment results on the Tecnick\cite{tecnick} dataset.}  
\label{fig:result_tecnick}  

\end{figure}  

\section{Comparative experiments with more methods}
To better demonstrate the performance of our method, we extend the experiments to include additional baseline approaches. These include Learned Image Compression methods WeConvene\cite{ECCV-2024-weconvene}, and FTIC\cite{2024-ICLR-FTIC}
and Perceptual methods MRIC\cite{agustsson2023multi}, TACO\cite{ICML-2024-TACO}. 
As shown in Fig.2, UGDiff achieves higher PSNR than MRIC and MS-ILLM on CLIC2020 dataset across all bitrates, demonstrating its strength in preserving pixel-level accuracy. Compared with WeConvene and FTIC, which mainly focus on improving structural consistency, UGDiff exhibits only a minor PSNR performance gap while achieves an obvious LPIPS performance improvment. TACO's superior LPIPS performance comes from its text-guided encoding strategy, which may not be universally applicable or desirable for all use cases. While MS-ILLM achieve extremely low LPIPS values, this often comes at the cost of structural fidelity (as seen in their lower PSNR compared to UGDiff). Our experiments results confirm that UGDiff achieves a superior balance between structural fidelity (PSNR) and perceptual quality (LPIPS) compared to the SOTA methods. This validates our proposed paradigm of wavelet conditional diffusion prediction followed by uncertainty-guided residual compression.

\begin{figure}[h!]  
\centering  
\includegraphics[width=1.0\linewidth]{figs/clic2020_psnr_lpips_add.pdf} 
\vspace{-5mm}
\caption{Quantitative comparisons with more methods in terms of perceptual quality (PSNR$\uparrow$ / LPIPS$\downarrow$/ on CLIC2020~\cite{clic2020} datasets.}  
\label{fig:result_clic2020}  

\end{figure}

\bibliographystyle{IEEEtran}
\bibliography{main}